\documentclass[preprint,prd,amsmath,amssymb,floatfix,nofootinbib]{revtex4}
\usepackage[colorlinks=true,urlcolor=black]{hyperref}

\hypersetup{backref = true,pagebackref = true,hyperindex = true, colorlinks = true,
  breaklinks = true, urlcolor = blue, linkcolor = blue, bookmarks = true,
  bookmarksopen = true, citecolor=red}

\usepackage{bm}
\usepackage{color}
\usepackage{dcolumn}
\usepackage{graphics}
\usepackage{graphicx}
\usepackage{subfigure}
\usepackage{slashed}
\usepackage{placeins}

\allowdisplaybreaks
\usepackage{pdfpages}
\usepackage{eso-pic}
\usepackage{everyshi}
\usepackage{pdflscape}

\newcommand{\bea}{\begin{eqnarray}}
\newcommand{\eea}{\end{eqnarray}}
\newcommand{\be}{\begin{equation}}
\newcommand{\ee}{\end{equation}}

\newcommand{\ar}{a_s}

\begin{document}

\title{Bjorken sum rule with analytic coupling at low $Q^2$ values}
\author{I.R. Gabdrakhmanov$^{1}$, N.A Gramotkov$^{1}$, A.V.~Kotikov$^{1}$, D.A. Volkova$^{1,2}$ and I.A.~Zemlyakov$^{1,3}$}
\affiliation{$^1$
  Joint Institute for Nuclear Research, 141980 Dubna, Russia;\\
 $^2$Dubna State University,
  141980 Dubna, Moscow Region, Russia;\\
  $^3$Tomsk State University,
  634010 Tomsk,
  Russia
}

\date{\today}

\begin{abstract}

  The experimental data obtained for the polarized Bjorken sum rule $\Gamma^{p-n}_1(Q^2)$ for small values of $Q^2$
  are approximated by the predictions obtained in the framework of analytic QCD up to the 5th order perturbation theory, whose coupling constant
  does not contain the Landau pole.
  We found an excellent agreement between the experimental data and the predictions of analytic QCD, as well as a strong difference between
  these data and the results obtained in the framework of
  perturbative
  QCD.
 
\end{abstract}

\maketitle

\section{Introduction}

Polarized Bjorken sum rule (BSR) $\Gamma^{p-n}_1(Q^2)$ \cite{Bjorken:1966jh,Bjorken:1969mm}, i.e. the difference between the first moments of the spin-dependent
structure functions (SFs) of a proton and neutron, is a very important space-like QCD observable \cite{Deur:2018roz,Kuhn:2008sy}.
Its isovector nature facilitates its theoretical description in perturbative QCD (pQCD) in terms of the operator product expansion (OPE),
compared to the corresponding SF integrals for each nucleon.
Experimental results for this quantity obtained in polarized deep inelastic scattering (DIS) are currently available in a wide
range of the spacelike squared momenta $Q^2$: 0.021 GeV$^2 \leq Q^2 <$ 5 GeV$^2$
\cite{Deur:2021klh,E143:1998hbs,E154:1997xfa,E142:1996thl,E155:1999pwm,E155:2000qdr,SpinMuon:1993gcv,SpinMuonSMC:1997voo,SpinMuonSMC:1994met,SpinMuon:1995svc,SpinMuonSMC:1997ixm,SpinMuonSMC:1997mkb,COMPASS:2005xxc,Compass:2007qxf,COMPASS:2010wkz,COMPASS:2015mhb,COMPASS:2016jwv,COMPASS:2017hef,HERMES:1997hjr,HERMES:1998cbu,HERMES:2006jyl,Deur:2004ti,Deur:2008ej,Deur:2014vea,ResonanceSpinStructure:2008ceg}.
In particular, the most recent experimental results \cite{Deur:2021klh} with significantly
reduced statistical uncertainties, derived mainly from the Jefferson Lab EG4 experiment on
polarized protons and deuterons and E97110 one on polarized ${}^3H$,
make BSR an attractive quantity for testing various pQCD generalizations at low $Q^2$ values: $Q^2 \leq 1$ GeV$^2$.

Theoretically, pQCD (with OPE) in the $\overline{MS}$-scheme was the usual approach to describing such quantities.
This approach, however, has the theoretical disadvantage that the running coupling constant ({\it couplant}) $\alpha_s(Q^2)$
has the Landau singularities for small $Q^2$ values: $Q^2 \leq 0.1$ GeV$^2$, which makes it inconvenient for estimating spacelike observables at small $Q^2$,
such as BSR.
In the recent years, the extension of pQCD couplings for low $Q^2$ without Landau singularities called (fractional) analytic perturbation theory [(F)APT)]
\cite{ShS,Shirkov:1997wi,MSS,Shirkov:2000qv,Shirkov:2001sm,BMS1,Bakulev:2006ex,Bakulev:2010gm}
(or the minimal analytic (MA) theory \cite{Cvetic:2008bn}), were applied to match the theoretical OPE expression with the experimental BSR data
\cite{Pasechnik:2008th,Pasechnik:2009yc,Kotikov:2012eq,Khandramai:2011zd,Ayala:2017uzx,Ayala:2017ucf,Ayala:2018ulm,Ayala:2020scz}.

Following \cite{Cvetic:2006mk,Cvetic:2006gc}, we introduce here the derivatives (in the $k$-order of perturbation theory (PT))
\be
\tilde{a}^{(k)}_{n+1}(Q^2)=\frac{(-1)^n}{n!} \, \frac{d^n a^{(k)}_s(Q^2)}{(dL)^n},~~a^{(k)}_s(Q^2)=\frac{\beta_0 \alpha^{(k)}_s(Q^2)}{4\pi}=\beta_0\,\overline{a}^{(k)}_s(Q^2),
\label{tan+1}
\ee
which are very convenient in the case of analytic QCD. $\beta_0$ is the first coefficient of the QCD $\beta$-function:
\be
\beta(\overline{a}^{(k)}_s)=-{\left(\overline{a}^{(k)}_s\right)}^2 \bigl(\beta_0 + \sum_{i=1}^k \beta_i {\left(\overline{a}^{(k)}_s\right)}^i\bigr),
\label{bQCD}
\ee
where $\beta_i$ are known up to $k=4$ \cite{Baikov:2008jh}.

The series of derivatives $\tilde{a}_{n}(Q^2)$ can successfully replace the corresponding series of $\ar$-powers (see, e.g. \cite{Kotikov:2022swl}). Indeed, each
derivative reduces the $\ar$ power but is accompanied by an additional $\beta$-function $\sim \ar^2$.
Thus, each application of a derivative yields an additional $\ar$, and thus  it is indeed possible to use a series of derivatives instead of
a series of $\ar$-powers.

In LO, the series of derivatives $\tilde{a}_{n}(Q^2)$ are exactly the same as $\ar^{n}$. Beyond LO, the relationship between $\tilde{a}_{n}(Q^2)$
and $\ar^{n}$ was established in \cite{Cvetic:2006mk,Cvetic:2010di} and extended to the fractional case, where $n \to$ is a non-integer $\nu $, in Ref.
\cite{GCAK}.

In this short paper, we apply the inverse logarithmic expansion of the MA couplants, recently obtained in \cite{Kotikov:2022sos,Kotikov:2023meh} for any PT order
(see Ref. \cite{Kotikov:2022vnx,Kotikov:2023nvz} for a brief introduction).
This approach is very convenient: for LO the MA couplants have simple representations (see \cite{BMS1}), while beyond LO the MA couplants are very close to LO ones,
especially for $Q^2 \to \infty$ and $Q^2 \to 0$, where the differences between MA couplants of various PT orders become insignificant.
Moreover, for $Q^2 \to \infty$ and $Q^2 \to 0$ the (fractional) derivatives of the MA couplants with $n\geq 2$ tend to zero, and therefore only the first
term in perturbative expansions makes a valuable contribution.

\section{Bjorken sum rule}

The polarized BSR
is defined as the difference between the proton and neutron polarized SFs,
integrated over the entire interval $x$
\be
\Gamma_1^{p-n}(Q^2)=\int_0^1 \, dx\, \bigl[g_1^{p}(x,Q^2)-g_1^{n}(x,Q^2)\bigr].
\label{Gpn} 
\ee

Theoretically, the quantity can be written in the OPE
form
(see Ref. \cite{Shuryak:1981pi,Balitsky:1989jb})
\be
\Gamma_1^{p-n}(Q^2)=
\frac{g_A}{6} \, \bigl(1-D_{\rm BS}(Q^2)\bigr) + \sum_{i=2}^{\infty} \frac{\mu_{2i}(Q^2)}{Q^{2i-2}} \, ,
\label{Gpn.OPE} 
\ee
where $g_A$=1.2762 $\pm$ 0.0005 \cite{PDG20} is
the nucleon axial charge, $(1-D_{BS}(Q^2))$ is the leading-twist
contribution, and $\mu_{2i}/Q^{2i-2}$ $(i\geq 1)$ are the higher-twist (HT)
contributions.
\footnote{Below, in our analysis, the so-called elastic contribution will always be excluded.}

Since we include very small $Q^2$ values here, the representation (\ref{Gpn.OPE}) of the HT contributions is inconvenient.
It is much better to use the so-called ``massive'' representation for the HT part (introduced in Ref. \cite{Teryaev:2013qba,Khandramai:2016kbh}):
\be
\Gamma_1^{p-n}(Q^2)=
\frac{g_A}{6} \, \bigl(1-D_{\rm BS}(Q^2)\bigr) +\frac{\hat{\mu}_4 M^2}{Q^{2}+M^2} \, ,
\label{Gpn.mOPE} 
\ee
where the values of $\hat{\mu}_4$ and $M^2$ have been fitted in Refs. \cite{Ayala:2017uzx,Ayala:2018ulm}
in the different analytic QCD models.

In the case of MA QCD, from \cite{Ayala:2018ulm} one can see that in (\ref{Gpn.mOPE})
\be
M^2=0.439 \pm 0.012 \pm 0.463
~~\hat{\mu}_{\rm{MA},4}
=-0.173 \pm 0.002\pm 0.666\,,
\label{M,mu} 
\ee
where
the statistical (small) and systematic (large)
uncertainties are presented.

Another form, which is correct at very small $Q^2$ values, has been proposed in \cite{Gabdrakhmanov:2017dvg}
\be
\Gamma_1^{p-n}(Q^2)=
\frac{g_A}{6} \, \bigl(1-D_{\rm BS}(Q^2)\bigr) +\frac{\hat{\mu}_4M^2(Q^{2}+M^2)}{(Q^{2}+M^2)^2+M^2\sigma^2} \, ,
\label{Gpn.mOPE.T} 
\ee
where small value $\sigma \equiv \sigma_{\rho}  =145$ MeV (the $\rho$-meson decay width) has been used.

Up to the $k$-th PT order,
the perturbative part has the form
\be
D^{(1)}_{\rm BS}(Q^2)=\frac{4}{\beta_0} \, a^{(1)}_s,~~D^{(k\geq2)}_{\rm BS}(Q^2)=\frac{4}{\beta_0} \, a^{(k)}_s\left(1+\sum_{m=1}^{k-1} d_m \bigl(a^{(k)}_s\bigr)^m
\right)\,,
\label{DBS} 
\ee
where $d_1$, $d_2$ and $d_3$ are known from exact calculations (see, e.g., \cite{Chen:2006tw,Chen:2005tda}). The exact $d_4$ value is not known, but it was recently
estimated  in Ref. \cite{Ayala:2022mgz}.

Converting the powers of couplant into its derivatives, we have
\be
D^{(1)}_{\rm BS}(Q^2)=\frac{4}{\beta_0} \, \tilde{a}^{(1)}_1,~~D^{(k\geq2)}_{\rm BS}(Q^2)=
\frac{4}{\beta_0} \, \left(\tilde{a}^{(k)}_{1}+\sum_{m=2}^k\tilde{d}_{m-1}\tilde{a}^{(k)}_{m}
\right),
\label{DBS.1} 
\ee
where
\bea
&&\tilde{d}_1=d_1,~~\tilde{d}_2=d_2-b_1d_1,~~\tilde{d}_3=d_3-\frac{5}{2}b_1d_2-\bigl(b_2-\frac{5}{2}b^2_1\bigr)\,d_1,\nonumber \\
&&\tilde{d}_4=d_4-\frac{13}{3}b_1d_3 -\bigl(3b_2-\frac{28}{3}b^2_1\bigr)\,d_2-\bigl(b_3-\frac{22}{3}b_1b_2+\frac{28}{3}b^3_1\bigr)\,d_1
\label{tdi} 
\eea
and $b_i=\beta_i/\beta_0^{i+1}$.

For the case of 3 active quark flavors ($f=3$), we have
\footnote{The coefficients $\beta_i$ $(i\geq 0)$ of the QCD $\beta$-function (\ref{bQCD})
  and, consequently, the couplant $\alpha_s(Q^2)$ itself depend on  the number $f$ of active quark flavors, and each new quark enters/leaves
  the game at a certain threshold $Q^2_f$ according to \cite{Chetyrkin:2005ia,Schroder:2005hy,Kniehl:2006bg}. The corresponding QCD parameters
  $\Lambda^{(f)}$ in N$^i$LO of PT can be found
  in Ref. \cite{Chen:2021tjz}.}
\bea
&&d_1=1.59,~~d_2=3.99,~~d_3=15.42~~d_4=63.76, \nonumber \\ 
&&\tilde{d}_1=1.59,~~\tilde{d}_2=2.73,
~~\tilde{d}_3=8.61,~~\tilde{d}_4=21.52 \, ,
\label{td123} 
\eea
i.e., the coefficients in the series of derivatives are slightly smaller.

In MA QCD, the results (\ref{Gpn.mOPE.T}) become as follows
\be
\Gamma_{\rm{MA},1}^{p-n}(Q^2)=
\frac{g_A}{6} \, \bigl(1-D_{\rm{MA,BS}}(Q^2)\bigr) +\frac{\hat{\mu}_{\rm{MA},4}M^2(Q^{2}+M^2)}{(Q^{2}+M^2)^2+M^2\sigma^2},~~
\,,
\label{Gpn.MA} 
\ee
where the perturbative part $D_{\rm{BS,MA}}(Q^2)$
takes the form
\be
D^{(1)}_{\rm MA,BS}(Q^2)=\frac{4}{\beta_0} \, A_{\rm MA}^{(1)},~~
D^{k\geq2}_{\rm{MA,BS}}(Q^2) =\frac{4}{\beta_0} \, \Bigl(A^{(k)}_{\rm MA}
+ \sum_{m=2}^{k} \, \tilde{d}_{m-1} \, \tilde{A}^{(k)}_{\rm MA,\nu=m} \Bigr)\,.
\label{DBS.ma} 
\ee

\section{Results}

\begin{table}[t]
\begin{center}
\begin{tabular}{|c|c|c|c|}
\hline
& $M^2$ for $\sigma=\sigma_{\rho}$ & $\hat{\mu}_{\rm{MA},4}$  for $\sigma=\sigma_{\rho}$ & $\chi^2/({\rm d.o.f.})$ for $\sigma=\sigma_{\rho}$ \\
& (for $\sigma=0$) & (for $\sigma=0$) & (for $\sigma=0$) \\
 \hline
 LO & 1.592 $\pm$ 0.300 & -0.168 $\pm$ 0.002 & 0.788  \\
 & (1.631 $\pm$ 0.301) & (-0.166 $\pm$ 0.001) & (0.789)  \\
 \hline
 NLO & 1.505 $\pm$ 0.286 & -0.157 $\pm$ 0.002 & 0.755  \\
 & (1.545 $\pm$ 0.287) & (-0.155 $\pm$ 0.001) & (0.757)  \\
 \hline
  N$^2$LO & 1.378 $\pm$ 0.242 & -0.159 $\pm$ 0.002 & 0.728  \\
 & (1.417 $\pm$ 0.241) & (-0.156 $\pm$ 0.002) & (0.728)  \\
 \hline
  N$^3$LO & 1.389 $\pm$ 0.247 & -0.159 $\pm$ 0.002 & 0.747  \\
  & (1.429 $\pm$ 0.248) & (-0.157 $\pm$ 0.002) & (0.747)  \\
   \hline
  N$^4$LO & 1.422 $\pm$ 0.259 & -0.159 $\pm$ 0.002 & 0.754  \\
 & (1.462 $\pm$ 0.259) & (-0.157 $\pm$ 0.001) & (0.754)  \\
 \hline
\end{tabular}
\end{center}
\caption{%
The values of the fit parameters with $\sigma=\sigma_{\rho}$ ($\sigma=0$).}
\label{Tab:BSR}
\end{table}

\begin{figure}[!htb]
\centering
\includegraphics[width=0.98\textwidth]{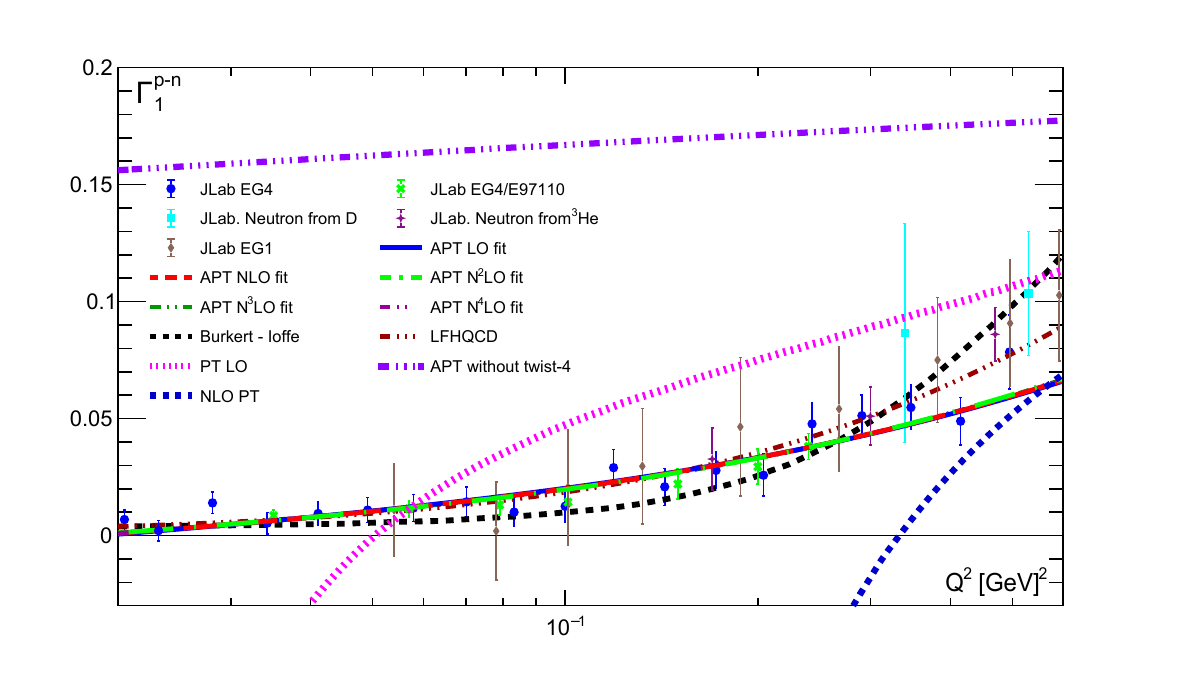}
\caption{\label{fig:APTHT}
  The results for $\Gamma_1^{p-n}(Q^2)$ in the first four orders of APT
  with $\sigma=\sigma_{\rho}$.
    }
\end{figure}

The calculation results taking into account only statistical uncertainties
are presented in Table \ref{Tab:BSR} and in Fig. \ref{fig:APTHT}.
Here we use the $Q^2$-independent $M$ and $\hat{\mu}_4$ values and the twist-two parts shown in Eqs. (\ref{DBS.1}) and (\ref{DBS.ma}) for the cases
of usual PT
and APT, respectively.

In the case of using MA couplants,  we see in Table \ref{Tab:BSR} that the cases $\sigma=0$ and $\sigma=\sigma_{\rho}$ lead to very similar values for
the fitting parameters and  $\chi^2$-factor.
So, in Fig. \ref{fig:APTHT} we show only the case
with $\sigma=\sigma_{\rho}$.
%
The quality of the fits is very good, as evidenced quantitatively by the values of $\chi^2/({\rm d.o.f.})$. Moreover, our results obtained for different PT orders
are very similar to each others: the corresponding curves in Fig. \ref{fig:APTHT}  are indistinguishable. One can also see the important role of the
twist-four term
(see also Refs. \cite{Khandramai:2011zd} and \cite{Kataev:2005ci,Kataev:2005hv} and discussions therein).
Without it, the value of $\Gamma_1^{p-n}(Q^2)$ is about 0.16, which is very far from the experimental data.

At $Q^2 \leq 0.3$ GeV$^2$
we also see good agreement with the phenomenological models: Burkert-Ioffe one \cite{Burkert:1992tg,Burkert:1993ya} and especially
LFHQCD one \cite{Brodsky:2014yha}. For larger $Q^2$ values our results are below the results of the phenomenological models
and at $Q^2 \geq 0.5$ GeV$^2$ are below the experimental data. We hope to improve agreement with using ``massive'' forms of HT
contributions $h_{2i}$ with $i\geq3$. This is a subject of future investigations.

As seen in Fig. \ref{fig:APTHT}, the results obtained using conventional couplants are not good  and worse for the NLO case to compare to the LO one. Indeed,
the deterioration increases with the PT order in this case (see \cite{Pasechnik:2008th,Pasechnik:2009yc,Ayala:2017uzx,Ayala:2018ulm,Kotikov:2022sos}).
Thus, the use of the ``massive'' twist-four form (\ref{Gpn.mOPE}) does not improve these results, since at $Q^2 \to \Lambda_i^2$ conventional
couplants become to be singular, that leads to large and negative results for the twist-two part (\ref{DBS}). As the PT order increases,
usual couplants become singular for ever larger $Q^2$ values, while BSR tends to negative values for ever larger $Q^2$ values
(see, e.g.,  Fig. 15 in \cite{Kotikov:2022sos}).
Thus, the discrepancy between theory and experiment  increases with the PT order.

\section{Conclusions}

We have considered the Bjorken sum rule in the framework of MA and
perturbative
QCD and obtained results similar to those obtained in previous
studies \cite{Pasechnik:2008th,Pasechnik:2009yc,Ayala:2017uzx,Ayala:2018ulm,Kotikov:2022sos} for the first 4 orders of PT.
The results based on the conventional PT do not agree with the experimental data. For some $Q^2$ values, the PT results become negative, since the
high-order corrections are large and enter the twist-two term with a minus sign.
APT in the minimal version leads to a good agreement with experimental data when we used the ``massive'' version (\ref{Gpn.MA}) for
the twist-four contributions.

Now we would like to discuss the photoproduction (PhP) case, i.e. the $Q^2\to0$ limit.
  In MA QCD,
  $A_{\rm MA}(Q^2=0)=1$ and $\tilde{A}^{(k)}_{\rm MA,m}=0$ for $m>1$ and we have
  \be
D_{\rm MA,BS}(Q^2=0)=\frac{4}{\beta_0} 
~~\mbox{and}~~
\Gamma_{\rm{MA},1}^{p-n}(Q^2=0,\sigma=0)=
\frac{g_A}{6} \, \bigl(1-\frac{4}{\beta_0}\bigr) +\hat{\mu}_{\rm{MA},4}
\,.
\label{Gpn.MA.Q0} 
\ee
The finitness of cross-section in the real photon limit 
leads \cite{Teryaev:2013qba,Khandramai:2016kbh,Gabdrakhmanov:2017dvg}
\be
\Gamma_{\rm{MA},1}^{p-n}(Q^2=0)=0
~~\mbox{and, hence,}~~
\hat{\mu}^{\rm php}_{\rm{MA},4}=-\frac{g_A}{6} \, \bigl(1-\frac{4}{\beta_0}\bigr).
\label{mu.GDH} 
\ee
In the case of 3 active quarks, i.e. $f=3$, we have
\be 
\hat{\mu}^{\rm php}_{\rm{MA},4}=-0.118
~~\mbox{and, hence,}~~
|\hat{\mu}^{\rm php}_{\rm{MA},4}|< |\hat{\mu}_{\rm{MA},4}|,
\label{mu.GDH} 
\ee
shown in Table I.\\
  So, in our fits the finitness of cross-section in the real photon limit
  is violated.
  \footnote{
    Note that the results for $\hat{\mu}_{\rm{MA},4}$ were obtained taking into account only statistical uncertainties.
    With systematic uncertainties, the results for $\hat{\mu}^{\rm php}_{\rm{MA},4}$ and $\hat{\mu}_{\rm{MA},4}$
    are in full agreement with each other, but the quality of our analysis is greatly reduced.}


  This is a common situation that appears as a consequence of the use of analytic versions of QCD for the Bjorken sum rule
  (see, e.g., Ref. \cite{Ayala:2018ulm}). Note that our results for $\hat{\mu}_4$ shown in Table I are smaller than in \cite{Ayala:2018ulm}.\\
In our future investigations 
  we plan to improve this analysis by taking several ``massive'' twists by analogy with twist-four one shown in Eq. (\ref{Gpn.mOPE}).
  We hope that this will lead to better agrement with the real photon limit and with the studies in Ref. \cite{Soffer:1992ck,Soffer:2004ip,Pasechnik:2010fg}.\\

Authors are grateful to Alexandre P. Deur for information about new experimental data in Ref. \cite{Deur:2021klh} and discussions.
The authors are also grateful to Andrei Kataev, Nikolai Nikolaev and Oleg Teryaev for criticism, leading to a sharp
improvement in the quality of the paper.
This work was supported in part by the Foundation for the Advancement of Theoretical
Physics and Mathematics “BASIS”.



\end{document}